# Simple Matrix Elements with Dynamical Fermions

The High Energy Monte Carlo Grand Challenge[*]

We report on studies of simple matrix elements from simulations with two flavors of sea quarks, both staggered and Wilson. We show the decay constants of vector and pseudoscalar mesons. The effects of sea quarks are small. These simulations are done at relatively large lattice spacing compared to most quenched studies.

## 1. INTRODUCTION

The HEMCGC has recently completed a series of simulations of QCD with two flavors of dynamical fermions on $16^3 \times 32$ lattices at a gauge coupling $\beta = 5.6$ with dynamical staggered fermions[1] and dynamical Wilson fermions[2] at $\beta = 5.3$. Here we would like to focus on our calculations of matrix elements and the effects of sea quarks on them. Some of this work has also been published in Ref. [3].

We wish to update one aspect of the Wilson spectroscopy: at Lattice 92 we[4] reported discrepancies in hadron masses calculated using interpolating fields on different source timeslices of the lattice. These effects have largely gone away with more statistics; we believe that they were due to long simulation time correlations in the data which led to an underestimate of uncertainties.

In his summary talk Weingarten[5] remarked that our time correlations were so long that we probably only had ten uncorrelated lattices in our data set. This is a gross misrepresentation of our data. Our autocorrelations are so long we probably only have two uncorrelated lattices. We encourage anyone who is going to revisit Wilson spectroscopy at $\beta = 5.3$ to collect three to five times as much data as we did, to be certain that time correlations in the data are under control.

## 2. OUR VERSION OF TADPOLE IMPROVEMENT (NOVEMBER 1993)

The lattice is a UV regulator and changing from the lattice cutoff to a continuum regulator (like $\overline{MS}$) introduces a shift

$$\langle f|O^{cont}(\mu = 1/a)|i\rangle_{\overline{MS}} =$$

$$a^D(1 + A_O \alpha_s) + \ldots)\langle f|O^{latt}(a)|i\rangle f(m)$$
$$+O(a) + \ldots \qquad (1)$$

where $A_O = \frac{1}{4\pi}(C_{\overline{MS}} - C_{latt})$, $f(m)$ converts the lattice field renormalization to the continuum and the factor $a^D$ converts the dimensionless lattice number to its continuum result. The $O(a)$ corrections arise because the lattice operator is not "improved". The C's are calculable in perturbation theory. Lepage and Mackenzie[6] have proposed a method for achieving a more convergent perturbation expansion called tadpole improvement, which we have attempted to implement in our calculations. Unfortunately there does not appear to be universal agreement on how to do this for all operators. Our present implementation of tadpole improvement (not the one we have used in Refs. [2] and [3]) is as follows:

For massless quarks and local operators $A_O$ consists of a sum of two terms: $A_O = A_{PT} + A_\kappa$ where $A_{PT}$ are the one loop perturbative corrections computed by (for example) Groot, Hoek and Smit[7] or Martinelli and Zhang[8]. $A_\kappa$ absorbs the perturbative shift in $\kappa_c$, $A_\kappa = 1.3643$

$$\frac{1}{2\kappa_c} = 4(1 - A_\kappa \alpha_s) \qquad (2)$$

and is included while changing the "conventional" bilinear field renormalization from $2\kappa_c$ to $1/4$. For "point-split" operators $\bar\psi \Gamma U \psi$ tadpole improvement makes the substitution $U \to u_0(U/u_0)$ and one is supposed to expand perturbatively in $U/u_0$. The quantity $u_0 = (\frac{1}{3}Tr U_P)^{1/4} = 1 - 1.0472\alpha_s$ and so there is an additional contribution to $A_O = A_{PT} + 1.3643 - 1.0472$ for nonlocal operators regardless of spatial orientation.

The coupling constant is defined through the

plaquette

$$-\ln\langle\frac{1}{3}TrU_P\rangle = \\ 4.18879\alpha_s(3.41/a)[1-(1.185+0.070n_f)\alpha_s+ \\ O(\alpha_s^2)] \qquad (3)$$

and is run down to a scale $O(1/a)$. These are our conventions. We have already published the uncorrected lattice data (as well as non-tadpole-improved results) so that if the theoretical situation changes one can go back and recompute everything.

Finally, if we are not at zero mass, $f(m) = \sqrt{(1-0.75\kappa_1/\kappa_c)(1-0.75\kappa_2/\kappa_c)}$. This is the "exp(ma)" field renormalization factor of Bernard, Labrenz and Soni [9].

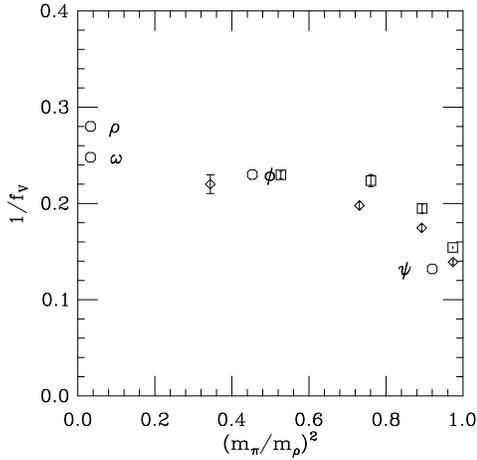

Figure 1. Sea quark mass dependence of vector meson decay constant.

## 3. SOME RESULTS

### 3.1. Vector currents

Fig. 1 shows $1/f_V$ the vector meson decay constant using the conserved (Wilson) current from our dynamical Wilson simulations (squares with $\kappa = 0.1670$, diamonds with $\kappa = 0.1675$). Fig. 2 shows three different lattice currents (crosses for the Wilson current, diamonds for the point-split current and squares for the local current) all at $\kappa = 0.1675$, all with tadpole improvement.

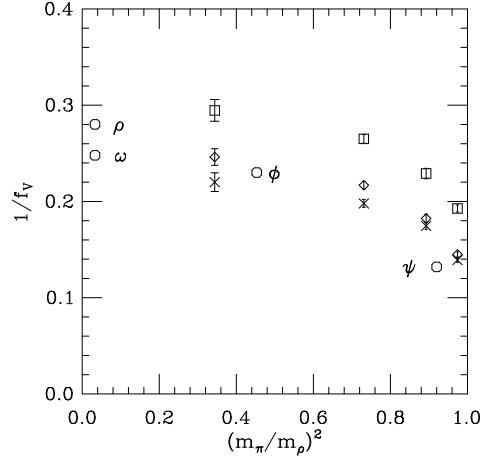

Figure 2. Operator dependence of vector meson decay constants.

The effects of sea quarks are much smaller than the systematic differences between different lattice choices for the operator.

### 3.2. Axial currents

Our extraction of pseudoscalar decay constants parallels other recent quenched analyses of these quantities. Note however that our lattice spacing is considerably larger than what is used in contemporary quenched simulations. This introduces an unknown systematic effect on our results.

In order to arrive at physical numbers we then carried out the following steps: First, we extrapolated heavy meson decay constants to zero light quark mass by a linear extrapolation in the light quark hopping parameter to $\kappa_c$, using the two lightest quark hopping parameters in each data set ($\kappa = 0.1615$ and either $0.1670$ or $0.1675$). This extrapolation included the $\kappa$-dependent field normalization and appropriate Z-factor. Our heavy quarks have masses which are large compared to an inverse lattice spacing. In this limit the dispersion relation for free Wilson fermions is

$$E(\vec{k}) = m_1 + \frac{\vec{k}^2}{2m_2} + \ldots \qquad (4)$$

where $m_1 = \log(\kappa_c/(2\kappa) - 3)$ and

$$am_2 = \frac{\exp(am_1)\sinh(am_1)}{1+\sinh(am_1)}. \qquad (5)$$

Bernard, Labrenz, and Soni [9], Kronfeld [10], and Mackenzie [11] argue that the appropriate quark mass at which the matrix element is measured is not $m_1$ but $m_2$ since it enters in the kinetic energy while $m_1$ is just an overall additive constant. Their analysis suggests that we correct for this error by adjusting the meson mass.

$$aM \to aM' = aM + (am_2 - am_1). \qquad (6)$$

This is a shift of no more than 0.125 at $\kappa = 0.1390$.

Fig. 3 displays a plot of $af_P\sqrt{aM_P}$ vs $1/aM_P$ for heavy-light systems, including the extrapolated zero light quark mass points. These are $\kappa = 0.1675$ simulations where crosses, diamonds, and squares show the local operators and octagons, bursts and fancy diamonds show the nonlocal operators. The fancy squares show the extrapolation to $\kappa_c$.

Finally, we convert to physical units by fixing the lattice spacing from $f_\pi$. We plot $f_P\sqrt{M_P}$ vs $1/M_P$ from our dynamical Wilson data in Fig. 4. Diamonds and octagons are local and nonlocal currents at $\kappa = 0.1670$; squares and crosses are the corresponding currents at $\kappa = 0.1675$.

We calculate $f_D = 215 \pm 40 \pm 35 \pm 5$ MeV and (from a long extrapolation to the B mass) $f_B = 150 \pm 40 \pm 40 \pm 5$ MeV. The three uncertainties represent lattice spacing, choice of operator (the local operators give about 250 MeV for $f_D$ and the nonlocal ones, about 180 MeV), and $\alpha_s$ uncertainty. The main sources of error are all systematic. Statistical errors for a calculation with a particular operator and choice of perturbative correction are never more than 5-10 MeV. We also find $f_{D_s} = 288 \pm 45 \pm 45 \pm 5$ MeV (The local operator gives 330 MeV, the nonlocal operator, 246 MeV.) We have included no uncertainty associated with the sea quark mass; it is lumped in with the statistical/extrapolation uncertainty. We do not see any observable effects of different sea quark masses. These calculations are a bit higher than quenched calculations done at smaller values of the lattice spacing [9,12].

It may be that the effect of dynamical fermions is to push up the matrix elements but it is also possible (and more likely, in our opinion) that the large lattice spacing induces a systematic shift upwards in the decay constant, especially for the tadpole-improved matrix elements. Fig. 5 shows a comparison of decay constants from our work and from the quenched simulations of Ref. [9], using the local axial current in both cases. The reader is invited to draw his/her/its own conclusions!

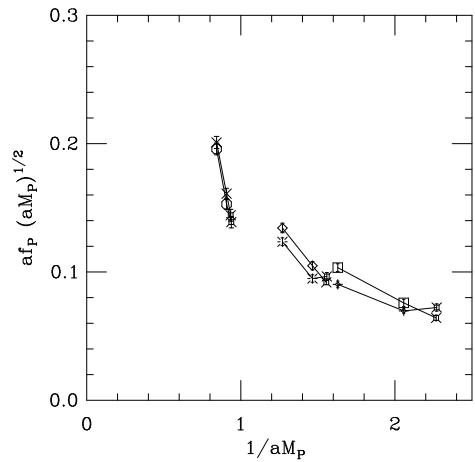

Figure 3. Lattice pseudoscalar decay constants.

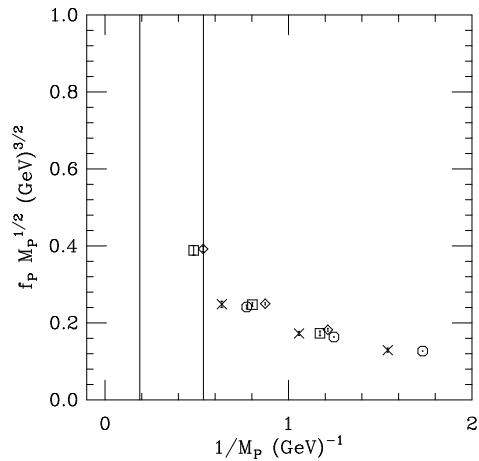

Figure 4. Pseudoscalar decay constants extrapolated to the continuum.

4

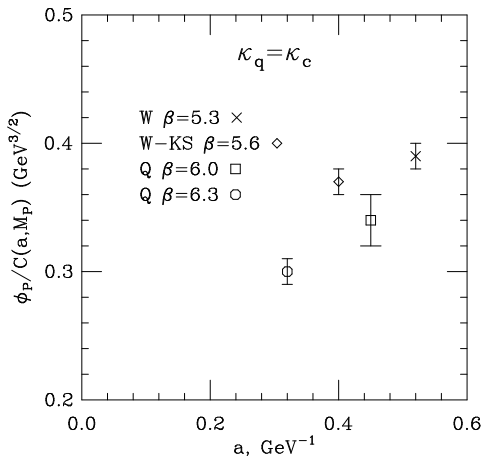

Figure 5. Quenched and dynamical decay constants vs. lattice spacing.

## 4. CONCLUSIONS

To really confront the question of whether sea quarks are important we need to push to smaller values of the sea quark mass. We also need either to push to smaller values of the lattice spacing or to continue to develop techniques which allow one to carry out simulations at large lattice spacing which have smaller intrinsic discretization systematics than present simulations do.

This work was supported by the U.S. Department of Energy and by the National Science Foundation. Simulations were performed at the Supercomputer Computations Research Institute.

## REFERENCES

[*] Khalil M. Bitar, R. Edwards, U. M. Heller and A. D. Kennedy, SCRI, Florida State University, Tallahassee, FL 32306-4052, USA; T. DeGrand, Physics Department, University of Colorado, Boulder, CO 80309, USA; Steven Gottlieb, Department of Physics, Indiana University, Bloomington, IN 47405, USA; J. B. Kogut, Physics Department, University of Illinois, Urbana, IL 61801, USA; A. Krasnitz, IPS, RZ F 3, ETH-Zentrum CH-8092 Zurich, SWITZERLAND; W. Liu, Thinking Machines Corporation, Cambridge, MA 02139, USA; Michael C. Ogilvie, Department of Physics, Washington University, St. Louis, MO 63130, USA; R. L. Renken, Physics Department, University of Central Florida, Orlando, FL 32816, USA; D. K. Sinclair, HEP Division, Argonne National Laboratory, 9700 South Cass Avenue, Argonne, IL 60439, USA; R. L. Sugar, Department of Physics, University of California, Santa Barbara, CA 93106, USA; D. Toussaint, Department of Physics, University of Arizona, Tucson, AZ 85721, USA; K. C. Wang, School of Physics, University of New South Wales, Kensington, NSW 2203, Australia. Poster presented by T. DeGrand.